\def\year{2021}\relax
\documentclass[letterpaper]{article} 
\usepackage{aaai21}  
\usepackage{times}  
\usepackage{helvet} 
\usepackage{courier}  
\usepackage[hyphens]{url}  
\usepackage{graphicx} 
\usepackage{natbib}  
\usepackage{caption} 
\usepackage[switch]{lineno}
\usepackage{xcolor}
\usepackage{bm}
\usepackage{amsmath}
\usepackage{latexsym}
\usepackage{array}
\usepackage{multirow}
\usepackage{color}
\definecolor{citecolor}{RGB}{0,0,0} 
\usepackage[pagebackref=false,breaklinks=true,letterpaper=true,colorlinks,citecolor=citecolor,bookmarks=false]{hyperref}%

\newcommand{\ie}{\emph{i.e.}}
\newcommand{\mL}{\mathcal{L}}

\def\D#1{{\color{black}#1}}

\title{Modeling the Probabilistic Distribution of Unlabeled Data for\\ One-shot Medical Image Segmentation}
\author {
        Yuhang Ding\textsuperscript{\rm 1},
        Xin Yu\textsuperscript{\rm 2},
        Yi Yang\textsuperscript{\rm 2}\thanks{Corresponding author} \\
}
\affiliations {
    \textsuperscript{\rm 1} Baidu Research, China \\
    \textsuperscript{\rm 2} ReLER, University of Technology Sydney, Australia \\
    dyh.ustc.uts@gmail.com, xin.yu@uts.edu.au, yi.yang@uts.edu.au
}
\begin{document}
\maketitle

\begin{abstract}
Existing image segmentation networks mainly leverage large-scale labeled datasets to attain high accuracy. However, labeling medical images is very expensive since it requires sophisticated expert knowledge. Thus, it is more desirable to employ only a few labeled data in pursuing high segmentation performance. 
In this paper, we develop \D{a data augmentation method for one-shot brain magnetic resonance imaging (MRI) image segmentation which exploits only one labeled MRI image (named atlas) and a few unlabeled images.}
In particular, we propose to learn the probability distributions of deformations (including shapes and intensities) of different unlabeled MRI images with respect to the atlas via 3D variational autoencoders (VAEs). In this manner, our method is able to exploit the learned distributions of image deformations to generate new authentic brain MRI images, and the number of generated samples will be sufficient to train a deep segmentation network. 
Furthermore, we introduce a new standard segmentation benchmark to evaluate the generalization performance of a segmentation network through a cross-dataset setting (collected from different sources).
Extensive experiments demonstrate that our method outperforms the state-of-the-art one-shot medical segmentation methods.
Our code has been released at \url{https://github.com/dyh127/Modeling-the-Probabilistic-Distribution-of-Unlabeled-Data}.

\end{abstract}

\section{Introduction}
Medical image segmentation aims to partition medical images, such as magnetic resonance imaging (MRI) image, into different anatomic regions. It plays an important role in many medical analysis applications, such as computer-assisted diagnosis and treatment planning.
In recent years, benefiting from deep convolution neural networks (CNNs), fully supervised medical image segmentation methods \cite{DBLP:conf/miccai/ZhouSTL18,DBLP:conf/cvpr/ChenWVCWZ19} have been extensively studied and achieved promising progress.
However, labeling anatomic regions for large-scale 3D images requires a huge amount of time and expert knowledge.
Hence, obtaining sufficient labelled data often becomes the bottleneck of fully supervised segmentation methods.

One-shot medical image segmentation, also called single atlas-based segmentation, has been proposed to reduce the demand for copious labeled data. 
Hand-crafted data augmentations \cite{DBLP:conf/miccai/RonnebergerFB15,DBLP:conf/3dim/MilletariNA16,DBLP:conf/miccai/RothLFSLTS15,DBLP:journals/tmi/PereiraPAS16}, such as random elastic deformations, generate new labeled images to improve segmentation performance. However, those methods often generate non-realistic images since they do not take the distribution of real images into account. Thus, their learned segmentation networks may not generalize well on real data.
Recently, deep learning based data augmentation methods \cite{DBLP:conf/cvpr/ZhaoBDGD19,DBLP:conf/miccai/XuN19,DBLP:conf/ipmi/ChaitanyaKBBDK19,DBLP:conf/cvpr/Wang0WWMWMZ20,DBLP:conf/wacv/ZhuMXLRHMX20} have been exploited. 
Those methods often leverage image registration to obtain profile and intensity differences between the atlas and other MR images, and then combine the profiles and intensities to generate new images for segmentation.
\begin{figure}[t]
\centering
\scalebox{1}[1]{\includegraphics[width=0.99\linewidth]{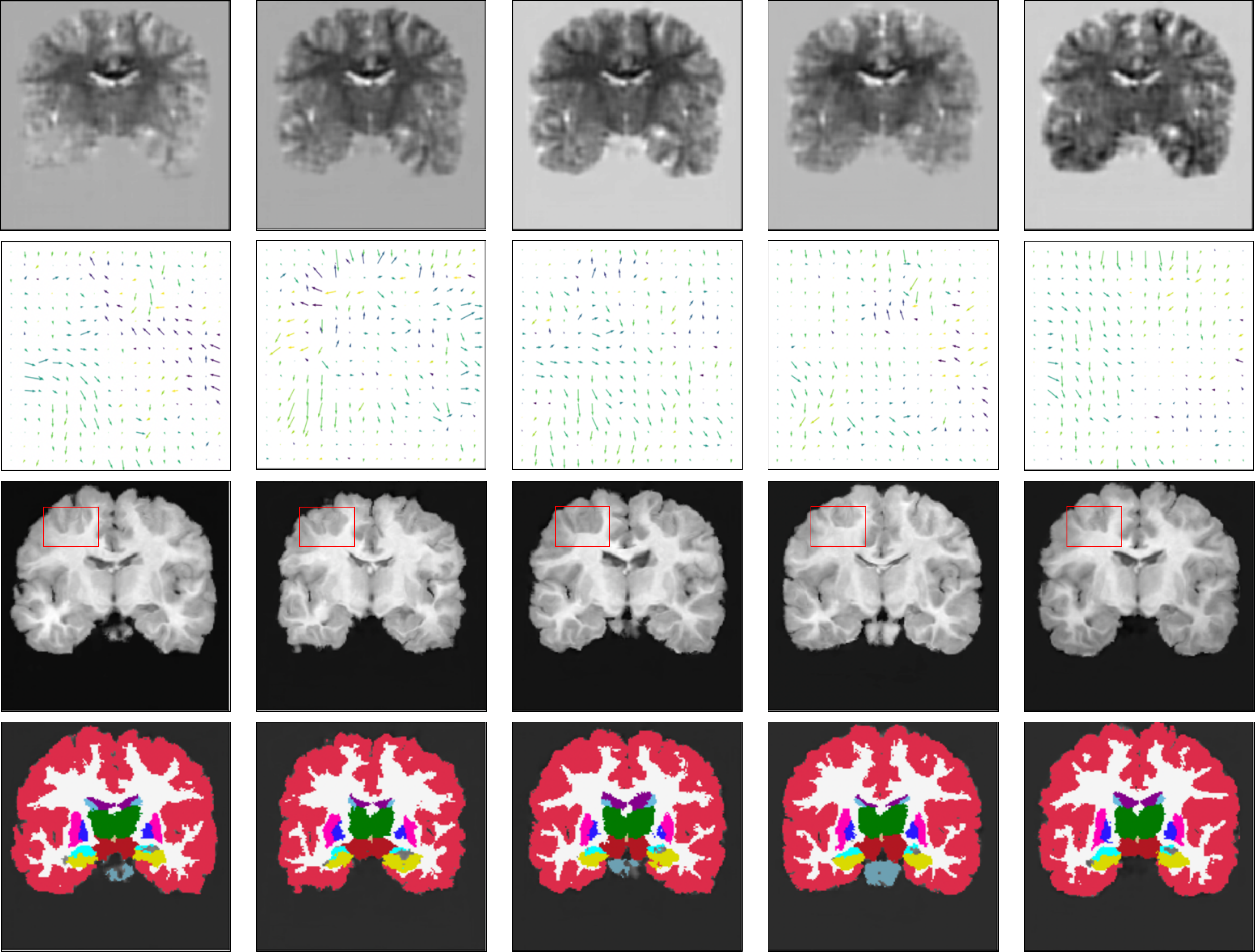}}
\vspace{-0.5em}
\caption{Illustration of our generated diverse deformations. 
From top to bottom: intensity offsets, shape deformations, synthesized images using the corresponding deformations and segmentation labels. Red frames highlight variations.
}\vspace{-1.0em}
\label{Fig:vae gen}
\end{figure}


Considering the domain gap and insufficient variations of synthesized data by previous methods, we aim to develop a novel medical image (\ie, MRI) augmentation method to address one-shot medical image segmentation tasks.
To this end, we propose probabilistic data augmentation approach to generate sufficient training images while ensuring them to follow the distribution of real MRI images in terms of brain shapes and MRI intensities, as shown in Fig~\ref{Fig:vae gen}. Thus, our segmentation network trained on our synthesized data will be robustly adapted to real MRI images.

In this work, we firstly employ image registration to obtain the shape deformations and intensity changes between an unlabeled MRI image and the atlas. However, since registration errors might occur in the registration procedure, directly classifying the registered images will lead to erroneous segmentation results. 
The prior art~\cite{DBLP:conf/cvpr/ZhaoBDGD19} combines the registered deformation fields and intensity changes to produce new images (with segmentation masks) and exploits them to train a segmentation network, thus mitigating registration errors. However, \citet{DBLP:conf/cvpr/ZhaoBDGD19} cannot provide new deformation fields and intensity changes. Therefore, the variety of generated images is still limited.

In contrast to prior works, we propose to exploit two variational autoencoders (VAEs) to capture the probabilistic distributions of deformation fields and intensity offsets with respect to the atlas. After that, our VAEs are employed to generate various profile deformations and intensity changes. The generative deformation fields and intensity variations are used to synthesize new MRI images. In this manner, our synthesized training data is not only abundant and diverse but also authentic to the real MRIs. 
Hence, using our augmented data, we improve the performance of our segmentation network significantly and achieve superior performance compared with the state-of-the-art.

Since different MRI machines (\ie, imaging sources) may lead to different characteristics in MRI images, such as intensity changes and signal-to-noise ratio, we also conduct experiments on unseen MRI sources to evaluate the robustness of our method. 
Thus, we propose a more challenging benchmark with an additional unseen test set.
Benefiting from our generated diverse training data, our segmentation network also performs better than the state-of-the-art on unseen MRI sources, thus demonstrating the superiority of our presented probabilistic augmentation method.


Overall, our contributions are threefold:
\begin{itemize}
    \item We propose a novel probabilistic data augmentation method based on VAEs to generate diverse and realistic training images for the downstream segmentation task. 
    \item We propose a new challenging segmentation benchmark to evaluate the performance of our proposed method and competing methods. It contains 3D brain MRI images from different sources. Thus, we can also test the generalization ability of the methods on unseen MRI sources.
    \item Taking advantage of our generated images, our method outperforms the state-of-the-art one-shot segmentation algorithms on both seen and unseen image sources.
\end{itemize}

\section{Related Work}
\subsection{Atlas-based Segmentation}
Atlas-based segmentation methods~\cite{DBLP:journals/bmcmi/KleinMGTH05,DBLP:journals/neuroimage/HeckemannHARH06} aim to segment target images by exploring knowledge from single or multiple labeled atlas images as well as a few unlabeled training images.
Because only a few labeled images are required, atlas-based segmentation methods are more desirable but challenging compared to fully supervised methods. 

Single atlas-based segmentation methods \cite{DBLP:conf/cvpr/Wang0WWMWMZ20,dinsdale2019spatial} leverage a registration model to learn shape deformations by aligning an atlas to target images, and then transfer the atlas label to the unlabeled ones as target labels. On the other hand, multi-atlas-based segmentation methods mainly focus on atlas selection \cite{DBLP:journals/mia/YangSLWX18} and label fusion \cite{DBLP:journals/mia/YangSLWX18,DBLP:conf/miccai/DingHN19}. Since our work belongs to the category of single atlas-based methods, we mainly review methods of this category as follows:

\citet{DBLP:conf/cvpr/Wang0WWMWMZ20} introduce a forward-backward consistency scheme into a registration network to obtain segmentation labels for unlabeled images. However, registration networks may suffer misalignment errors, thus leading to inferior segmentation results. 
Instead of directly transferring segmentation labels to a target image, some works have been proposed to warp the atlas image and its label to generate new images. Then, the generated data are used to train a segmentation network.
For example, \citet{DBLP:conf/cvpr/ZhaoBDGD19} leverage image registration to learn shape and intensity deformations between target images and the atlas. Then, new images synthesized by the learned deformations are exploited to train their segmentation network.
The works \cite{DBLP:conf/miccai/XuN19,DBLP:conf/wacv/ZhuMXLRHMX20} jointly learn image segmentation and registration simultaneously.
Since the deformations learned from the unlabeled data are deterministic and only a few, the diversity of generated images is limited.

In contrast, our proposed method is able to arbitrarily generate various shape and intensity deformations that even do not exhibit in any provided images. Moreover, our generated deformations are sampled from the distribution of the deformations between unlabeled images and the atlas via two VAEs. Hence, our synthesized MRI images are not only abundant but also authentic to real ones, thus facilitating the training of our segmentation network.

\subsection{Medical Image Data Augmentation}
Data augmentation is one of the most effective techniques for reducing over-fitting and improving the capability of networks.
In medical image analyses, a large number of data augmentation methods have been introduced due to the absence of large-scale labeled training data.

Traditional hand-crafted data augmentation methods are designed to deform medical images in terms of qualities, appearance or shapes.
\citet{DBLP:conf/miccai/ChristEETBBRAHD16} augment CT images with Gaussian noise while \citet{DBLP:journals/mia/Sirinukunwattana17} apply Gaussian blur to augment images for gland segmentation.
\citet{DBLP:conf/miua/DongYLMG17} enrich the training set by randomly enhancing the brightness of MRI images.
The works~\cite{DBLP:conf/miccai/RonnebergerFB15,DBLP:conf/miccai/CicekALBR16} exploit random elastic deformations to generate annotated images.

Generative adversarial networks (GANs) have been employed to synthesize new images \cite{DBLP:conf/miccai/MahapatraBT018,DBLP:conf/miccai/JinXTHM18,DBLP:conf/cvpr/FuLHHSDD18}. \citet{DBLP:conf/cvpr/FuLHHSDD18,DBLP:conf/aaai/Cao0WGS20} adopt CycleGANs~\cite{zhu2017unpaired} to generate 3D microscopy images while \citet{DBLP:conf/miccai/MahapatraBT018} use conditional GANs to synthesize realistic chest X-ray images. \citet{DBLP:conf/ipmi/ChaitanyaKBBDK19} exploit conditional GANs to generate deformations of brain MRI images. 
However, when only one labeled example and only a few unlabeled ones (less than 100 MRI images) are available, GAN based augmentation methods, in this case, might suffer mode collapse, such as outputting all zeros for shape and intensity deformations. 



\begin{figure*}[t]
\centering
\includegraphics[width=0.99\textwidth]{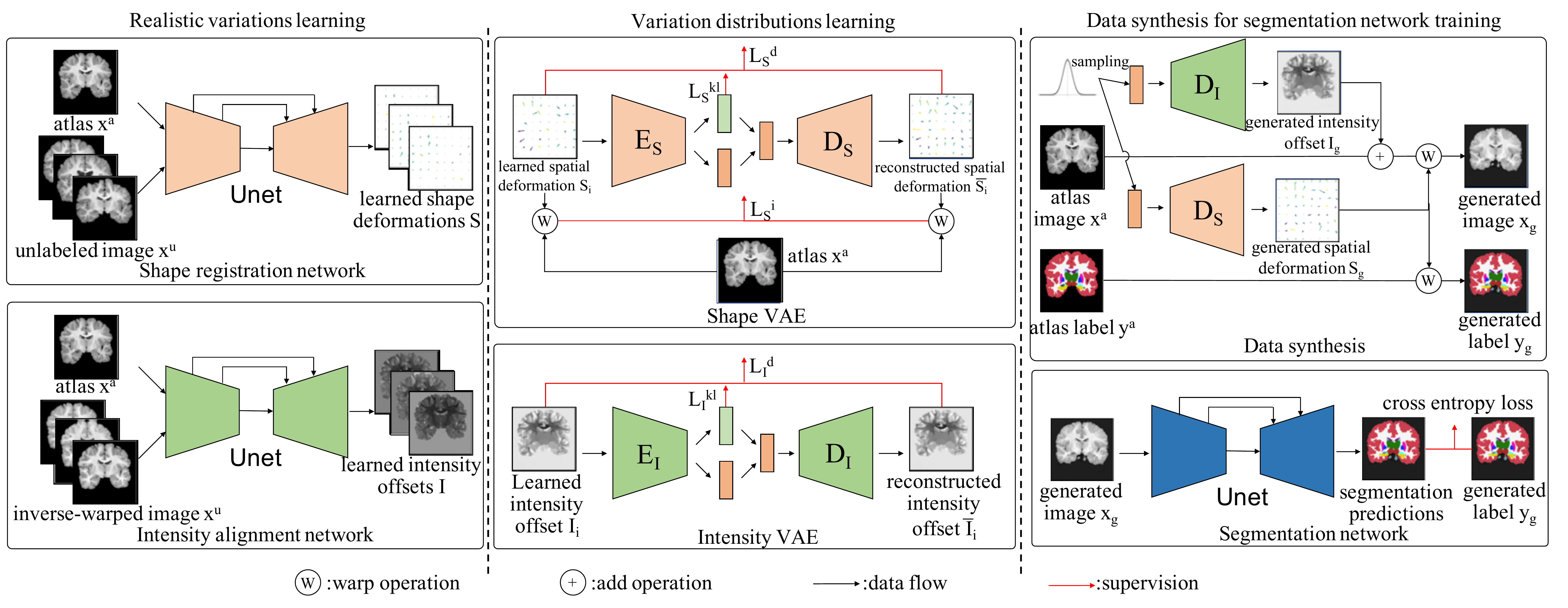}
\vspace{-1em}
\caption{The framework of our proposed method: (i) image deformations are obtained by two Unet-based registration networks; (ii) our shape and intensity VAEs are proposed to learn the variation distributions and generate new deformations; (iii) new training samples are synthesized by applying the generated deformations to the atlas image and our segmentation network is trained on these samples. 
}
\vspace{-1em}
\label{Fig:overview of methods}
\end{figure*}

\section{Proposed Method}
In this work, we leverage an image registration network and two VAEs to generate diverse and authentic brain MRI training samples. The generative samples are then employed to improve our segmentation network.
Here, we introduce the procedure of image registration as well as modeling the probabilistic distributions of those deformations via our shape and intensity 3D VAEs, respectively.

After obtaining the models of the deformations, we randomly sample from the distributions of the deformations and then construct new MRI images with the atlas image. The newly synthesized MRI images with their labels will be used to train our segmentation network.

\subsection{Learning Deformations from Image Registration}
Image registration \cite{DBLP:conf/aaai/MiaoPFTMML18,zitova2003image} aims to align an image to a template one, called atlas, by learning shape deformations between them. 
Most existing registration-based segmentation methods \cite{DBLP:conf/cvpr/Wang0WWMWMZ20,DBLP:conf/miccai/XuN19,DBLP:conf/wacv/ZhuMXLRHMX20} only consider the structure differences between two images.
However, due to different patients, scan machines and operations, image intensities also vary. Therefore, we model both shape and intensity deformations.

First, as shown in Fig.~\ref{Fig:overview of methods}, we leverage a Unet-based \cite{DBLP:conf/miccai/RonnebergerFB15} registration network (named shape registration network) to learn 3D shape deformations.
Denote an atlas image and its segmentation mask as ($x^a$, $y^a$) and $N$ unlabeled images as $\{x^u_1, x^u_2, \cdot \cdot \cdot, x^u_N\}$. Taking the atlas image $x^a$ and an unlabeled training image $x^u_i$ as the input, the registration network is trained to propagate the atlas image $x^a$ to an unlabeled image $x^u_i$ by estimating a shape deformation $S_i$.
In other words, $S_i$ is optimized to warp $x^a$ to $x^u_i$: $x^u_i \leftarrow x^a \circ S_i$, where $\circ$ represents a warping operation implemented by a differentiable bilinear interpolation-based spatial transformer layer \cite{DBLP:conf/cvpr/BalakrishnanZSG18}.
Following the work \cite{DBLP:journals/tmi/BalakrishnanZSG19}, we employ a local cross-correlation (CC) loss $\mL_{CC}$ and a deformation smoothness regularization $\mL_{S}^{reg}$ to train our shape registration network in an unsupervised manner and its objective $\mL_{srn}$ is formulated as:
\begin{equation}
    \begin{split}
         &\mL_{CC}\!=\!\sum_i \sum_{p \in \Omega} \frac{g(x_i^u, [x^a \circ S_i], p)^2}{g(x_i^u, x_i^u, p) g([x^a \circ S_i], [x^a \circ S_i], p)}, \\
         &\mL_{S}^{reg}\!=\!\sum\nolimits_i \Vert \nabla S_i \Vert_2, \\
         &\mL_{srn}\!=\!- \mL_{CC} + \mL_{S}^{reg},
           \end{split}
\end{equation}
where $g(a, b, p)$ denotes the correlation between local patches $a$ and $b$ on voxel $p$: $g(a, b, p) = \sum_{p_j} (a(p_j)-\overline{a}(p)) (b(p_j)-\overline{b}(p))$, and $\overline{a}(p)$ indicates the mean of local patch intensities on $p$: $\overline{a}(p) = \frac{1}{\Vert p \Vert_1} \sum_{p_j} a(p_j)$. 
$p$ represents a $n^3$ cube
in a 3D image $\Omega$ and $p_j$ denotes the pixels in the cube. We set $n$ to 9 similar to prior methods \cite{DBLP:journals/tmi/BalakrishnanZSG19}.
$\mL_{CC}$ encourages the structure similarities between two images regardless of the intensity variations while $\mL_{S}^{reg}$ aims to constrain shape deformations to be smooth.
$\nabla S_i$ denotes the spatial gradients of the shape variations.

Similar to learning shape deformations, we also use a Unet-based network, called intensity alignment network, to align 3D intensity deformations.
As visible in Fig.~\ref{Fig:overview of methods}, the network takes the atlas image $x^a$ and the inverse-warped image $\hat{x}^u_i$ as input to measure the intensity deformations $I_i$.
$\hat{x}^u_i$ is generated by aligning $x^u_i$ to $x^a$, and thus $\hat{x}^u_i$ and $x^a$ share similar profile structure.
Similar to \cite{DBLP:conf/cvpr/ZhaoBDGD19}, we exploit a pixel-wise reconstruction loss $\mL_{sim}$ between $x^a$ and $x^u_i$ and an intensity smoothness regularization $\mL_{I}^{reg}$ to train our intensity alignment network. The objective function $\mL_{irn}$ is expressed as:
\begin{equation}
    \begin{split}
         &\mL_{sim}\!=\!\sum\nolimits_i \Vert (x^a + I_i) \circ S_i - x^u_i \Vert_2, \\
         &\mL_{I}^{reg}\!=\!\sum\nolimits_i \sum\nolimits_{q_j} (1 - c^a(p_j)) |\nabla I_i(p_j)|, \\
         &\mL_{irn}\!=\! \mL_{sim} + \lambda \mL_{I}^{reg}.
           \end{split}
\end{equation}
Here, $\mL_{I}^{reg}$ is designed to prevent dramatic changes of the $I_i$ in the same brain area.
$\nabla I_i(p_j)$ denotes the gradients of $I_i$ at $p_j$.
$c^a$ denotes the mask of contours across different areas. 
$\lambda$ is a trade-off weight and set to 0.02, following the work~\cite{DBLP:conf/cvpr/ZhaoBDGD19}.

\subsection{Diverse Image Generation via VAEs}
After image registration, we obtain $N$ shape deformations and $N$ intensity changes from the atlas and $N$ unlabeled images.
In the work \cite{DBLP:conf/cvpr/ZhaoBDGD19}, these variations are directly combined to generate new labeled training images for segmentation.
However, only $N$ kinds of shape and intensity transformations are involved during training, and the diversity of the samples is not rich enough to train an accurate segmentation network. \citet{DBLP:conf/ipmi/ChaitanyaKBBDK19} employ GANs to generate new deformations but their method requires a large number of unlabeled data to train GANs. However, we only have less than 100 unlabeled images and their method will suffer mode collapse and is not applicable in our case.

Different from previous methods, we adopt a 3D shape VAE and a 3D intensity VAE to learn the probabilistic distributions of the variations with respect to the atlas separately, since VAE does not suffer mode collapse. Furthermore, inspired by beta-VAE~\cite{DBLP:conf/iclr/HigginsMPBGBML17,DBLP:journals/corr/abs-1804-03599}, we reduce impacts of the Kullback-Leibler (KL) divergence in a conventional VAE to increase the diversity of generated samples. Doing so is also driven by the insufficiency of the training samples. After training, we sample deformations from our shape and intensity VAEs, and then generate a large number of various training images.

As illustrated in Fig.~\ref{Fig:overview of methods}, our shape VAE first uses an encoder to project an input shape deformation into a latent vector $z = \boldsymbol{\mathrm{E}}_S(S_i)$ and then decodes $z$ to the image domain, \ie, a reconstructed shape deformation $\overline{S}_i = \boldsymbol{\mathrm{D}}_S(z)$.
During training, three objectives, including KL divergence $\mL_{S}^{kl}$ and pixel-wise reconstruction losses on the deformations $\mL_{S}^{d}$ and image intensities $\mL_{S}^{i}$, are employed to train our shape VAE, written as:
\begin{equation}
    \begin{split}
         &\mL_{S}^{kl} = \sum\nolimits_{i} D_{kl}(q(z|S_i)||p(z)), \\
         &\mL_{S}^{d} = \sum\nolimits_{i} \Vert S_i - \overline{S}_i \Vert_2, \\
         &\mL_{S}^{i} = \sum\nolimits_{i} \Vert (x^a \circ S_i) - (x^a \circ \overline{S}_i) \Vert_2, \\
         &\mL_{S} = (\mL_{S}^{d} + \mL_{S}^{i}) + \beta \mL_{S}^{kl}, \end{split}
         \label{eq:1}
\end{equation}
where $\mL_{S}^{kl}$ forces the distribution of latent vector $z$ to be a standard normal distribution, (\ie, $z \sim \mathcal{N}(0,1)$), $q(z|\cdot)$ denotes the posterior distribution, $p(z)$ denotes the Gaussian prior distribution modeled by a standard normal distribution, and $\beta$ is a hyper-parameter controlling rigidity of the distributions of the latent variable $z$ and the quality of reconstruction.
Here, we not only compare the decoded shape deformations with the input ones but also measure the differences between the warped images by the input shape deformations and reconstructed ones.

Smaller $\beta$ indicates less attention is paid to the KL divergence loss during training and will result in a larger KL divergence between the posterior and prior distributions.
As suggested by \citet{DBLP:journals/corr/abs-1804-03599}, larger KL divergence allows a latent vector to reside in a large space. In other words, smaller $\beta$ allows our VAE to preserve variations of input images especially when the training samples are scarce.
Therefore, using a small $\beta$ is more preferable when the number of training samples is limited.
Moreover, since the latent space has been enlarged, more variations can be generated from this latent vector space via our decoder in the testing phase.
Therefore, we set $\beta$ to a small value (\textit{i.e.,} $0.1$) for all the experiments.

It is worth noting that we employ both $\mL_{S}^{d}$ and $\mL_{S}^{i}$ as the reconstruction loss for our shape VAE instead of only reconstructing network inputs by $\mL_{S}^{d}$ as in the original VAE.
When $\mL_{S}^{d}$ is only employed, image structure information is neglected. 
In particular, shape deformations should pay attention to the consistency of image contour movements.
However, $\mL_{S}^{d}$ treats the movement of each pixel individually and thus may not perform consistent movements along the contour regions.
\D{On the contrary,} the reconstruction loss $\mL_{S}^{i}$ is sensitive to the movements of image contours because image intensities around contours change dramatically. In other words, small reconstruction errors in the deformations of the contours will lead to large intensity differences between two warped images.
\D{On the other hand, since $\mL_{S}^{i}$ only measures intensity similarities, it may not preserve boundary information when two areas have similar intensities.
}
Therefore, we leverage both $\mL_{S}^{i}$ and $\mL_{S}^{d}$ as the reconstruction loss in learning our shape VAE.

Similar to our shape VAE, we employ a VAE to model the distribution of the intensity variations with respect to the atlas. Here, we adopt the standard KL divergence loss and a pixel-wise reconstruction loss to train our intensity deformation VAE, expressed as:
\begin{equation}
    \begin{split}
         &\mL_{I}^{kl} = \sum\nolimits_i D_{kl}(q(z|I_i)||p(z)), \\
         &\mL_{I}^{d} = \sum\nolimits_{i} \Vert I_i - \overline{I}_i \Vert_2, \\
         &\mL_{I} = \mL_{I}^{d} + \beta \mL_{I}^{kl},
    \end{split}
    \label{eq:2}
\end{equation}
where $\overline{I}_i$ is the intensity deformation reconstructed from $I_i$.

After modeling the deformation distributions, our shape and intensity VAEs are exploited to generate diverse variations by random sampling.
Specifically, in the process of the generation, the decoders $\boldsymbol{\mathrm{D}}_S$ and $\boldsymbol{\mathrm{D}}_I$ take random latent vectors sampled from a Gaussian distribution $\mathcal{N}(0,\sigma)$ as input and output various shape deformations $S_g$ and intensity changes $I_g$, respectively. Then, our synthesized labeled training images are constructed as:
\begin{equation}
    x_g = (x^a + I_g) \circ S_g,~~~~~~~y_g = y^a \circ S_g,
\end{equation}
where $x_g$ and $y_g$ represent the synthesized images and their corresponding segmentation masks.
Note that, different from MRI images, segmentation masks are warped by a nearest-neighbor interpolation-based 3D spatial transformer layer \cite{DBLP:conf/cvpr/BalakrishnanZSG18}.

\subsection{Segmentation Network}
Once augmented training samples are obtained, we can train our segmentation network on those samples.
In order to conduct fair comparisons to the state-of-the-art \cite{DBLP:conf/cvpr/ZhaoBDGD19},
\D{we employ the same 2D Unet with a five-layer encoder and a five-layer decoder to segment each slice of 3D images individually.
In the encoder and decoder, we use 3x3 2D convolutional operations followed by LeakyReLU layers. 2x2 Max-pooling layers are used to reduce the feature resolution while upsampling layers are used to increase resolution by a factor of 2.}

In each training iteration, we randomly sample slices from 3D images to construct a batch.
The standard cross-entropy loss is employed as follows:
\begin{equation}
    \mL_{CE} = - \sum_{i=1}^W \sum_{j=1}^H \frac{1}{H \cdot W} \mathrm{log} \frac{\mathrm{exp}(y_p[i, j, y_g(i, j)])}{\sum_{k=1}^K \mathrm{exp} (y_p[i, j, k])},
\end{equation}
where $y_p$ is the predicted mask from our segmentation network g (\ie, $y_p = g(x_g; \theta)$) and $\theta$ denotes the parameters of the segmentation network.
$W$ and $H$ denote the width and height of a 2D slice, respectively.
$K$ indicates the number of anatomical components in an MRI image.
Similar to the training process, every 3D image is split into 2D slices and segmented in a slice-wise fashion in the testing phase.

Although we incorporate two VAEs to generate labeled data, they are only used in the training phase.
During testing, only our segmentation network is exploited.
Therefore, our method does not increase the network parameters and FLOPs during inference and thus can be deployed as easily as previous works.

\subsection{Implementation Details}
We adopt the same network architecture for our shape and intensity VAEs, and the VAEs are 3D VAEs since deformations should be consistent in 3D space. 
More details of the network architecture are described in the supplementary material.
In the 3D VAE networks, group normalization \cite{DBLP:journals/ijcv/WuH20} is employed. For the activation function, we use LeakyReLU and ReLU for the encoder and the decoder, respectively.
The dimension of the latent vector is set to $512$.

During training, Adam \cite{DBLP:journals/corr/KingmaB14} optimizer is used to train our VAEs, where $\beta_1$ and $\beta_2$ are set to $0.5$ and $0.999$, respectively. The batch size is set to $1$ due to the GPU memory limit.
The learning rate is fixed to $1e^{-4}$ for the whole $40k$ training iterations.
The hyper-parameter $\beta$ in both two VAEs is set to $0.1$.
In generating deformations, the shape VAE and the intensity VAE take latent vectors sampled from $\mathcal{N}(0,10)$ as input in order to achieve more diverse data.

For other networks (\ie, shape registration, intensity alignment and segmentation networks), a default Adam with $1e^{-4}$ learning rate is employed.
For the shape registration and intensity alignment networks, the batch size is set to 1 and the networks are trained for 500 epochs.
For the segmentation network, the batch size is set to $16$ and the network is trained for $40k$ iterations.
Our method is trained and tested on an Nvidia Tesla V100 GPU and achieves similar results on Keras with a TensorFlow backend and PaddlePaddle.

Note that, in training the 3D VAEs and segmentation networks, images are generated on-the-fly, and thus we train these networks in terms of iterations. In training registration and alignment networks, only 82 MRI images will be transformed to the atlas, and thus we train the networks in terms of epochs. 

\section{Experiments}
In this section, we first compare our proposed method with state-of-the-art one-shot based methods and then analyse the contributions of each component in our method. For fair comparisons, we conduct our experiments on the same dataset as previous works~\cite{DBLP:journals/tmi/BalakrishnanZSG19,DBLP:conf/cvpr/ZhaoBDGD19,DBLP:conf/cvpr/Wang0WWMWMZ20}.
Moreover, we propose a more challenging MRI benchmark
to evaluate the generalization performance of state-of-the-art one-shot based methods on unseen MRI data.




\subsection{Dataset and Evaluation Metric}
\textbf{Dataset:} CANDI dataset \cite{kennedy2012candishare} consists $103$ T1-weighted brain MRI images from $57$ males and $46$ females. 
In this dataset, four types of diagnostic groups are considered including healthy controls, schizophrenia spectrum, bipolar disorder with psychosis, and bipolar disorder without psychosis.
In the experiments, we use the same train and test splits as in \cite{DBLP:conf/cvpr/Wang0WWMWMZ20}. To be specific, $20$, $82$ and $1$ images are employed as the test set, unlabeled training set and atlas, respectively. 
Following the work~\cite{DBLP:conf/cvpr/Wang0WWMWMZ20}, we crop a $160 \times 160 \times 128$ volume from the center of an original MRI image.
For segmentation, similar to \cite{DBLP:conf/cvpr/Wang0WWMWMZ20}, we consider 28 primary brain anatomical areas.

\noindent\textbf{Evaluation Metric:} Dice coefficient \cite{dice1945measures} is used to measure the segmentation performance, written by:
\begin{equation}
    \mathrm{Dice}(M^k_{y_{p}}, M^k_{y_{gth}}) = 2 \cdot \frac{M^k_{y_{p}} \bigcap M^k_{y_{gth}}}{|M^k_{y_{p}}| + |M^k_{y_{gth}}|}, 
\end{equation}
where $M^k_{y_{p}}$ and $M^k_{y_{gth}}$ denote segmentation masks of the anatomical region $k$ with predicted labels $y_p$ and its corresponding ground-truth $y_{gth}$. 

Larger Dice scores indicate more overlaps between predictions and ground-truth labels, and thus represent better segmentation performance.
To better demonstrate the performance of each method, we report not only a mean Dice score but also its corresponding standard variance, minimum Dice score and maximum Dice score on the test set.

\subsection{Comparison with State-of-the-art Methods}
We mainly compare two state-of-the-art one-shot atlas based method, namely DataAug~\cite{DBLP:conf/cvpr/ZhaoBDGD19} and TL-Net~\cite{DBLP:conf/cvpr/Wang0WWMWMZ20}.
{In addition, one unsupervised registration method \ie, VoxelMorph~\cite{DBLP:conf/cvpr/BalakrishnanZSG18} is applied to one-shot medical image segmentation for comparison.}
VoxelMorph and TL-Net leverage a registration network to align the input MRI images to the atlas and then transfer the segmentation mask of the atlas to the input images as the segmentation results.
DataAug employs image registration to achieve shape and intensity transformation, and then augment the atlas image with the attained transformation to train a segmentation network. Note that these state-of-the-art methods do not generate new deformations while our method does.

\begin{figure*}[t]
\centering
\scalebox{1}[0.75]{\includegraphics[width=0.99\linewidth]{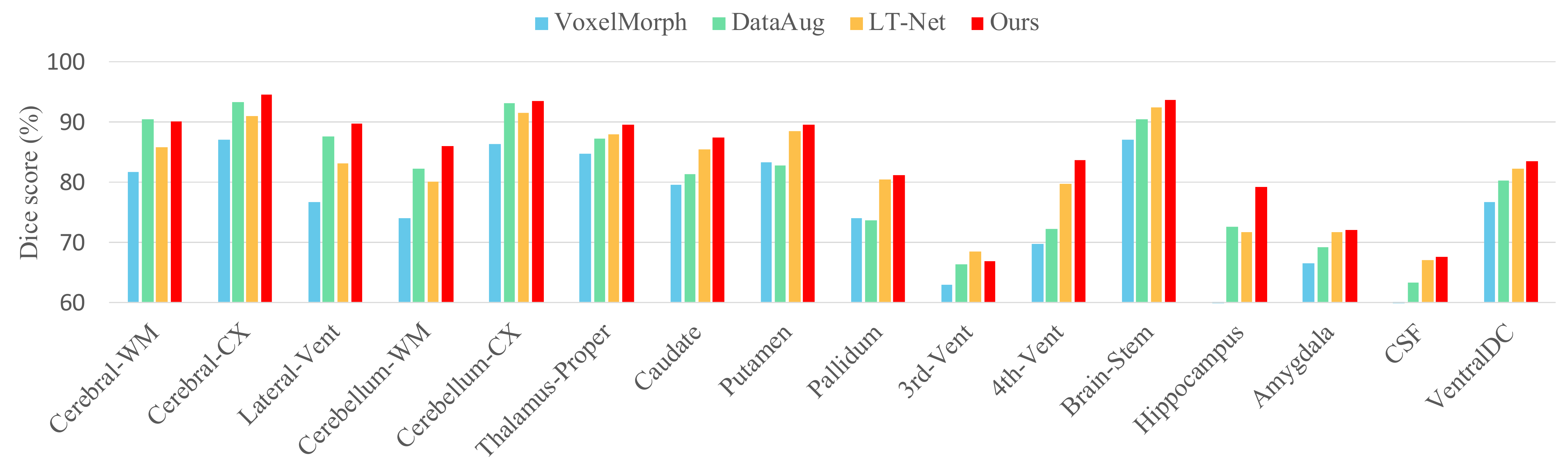}}
\vspace{-1em}
\caption{Comparison with the state-of-the-art on different brain areas. 
Left and right areas with the same labels are combined together.
The abbreviations WM, CX, vent, CSF denote white matter, cortex, ventricle and cerebrospinal fluid, respectively.
}
\vspace{-0.8em}
\label{Fig:CADNI SOTA}
\end{figure*}

As seen in Table \ref{tab:SOTA CANDI} and Fig.~\ref{Fig:CADNI SOTA}, we demonstrate the segmentation performance of our method is superior to that of the state-of-the-art.
As indicated in Table \ref{tab:SOTA CANDI}, our method achieves superior segmentation performance compared to the state-of-the-art. In particular, our method improves the segmentation performance by $2.8$\% on the Dice score in comparison to the second best method LT-Net \cite{DBLP:conf/cvpr/Wang0WWMWMZ20}. Moreover, our method also obtains the smallest variance, demonstrating that our method is more robust.
Figure~\ref{Fig:CADNI SOTA} demonstrates the segmentation results with respect to anatomical structure (symmetrical ones are reported together). As visible in Fig.~\ref{Fig:CADNI SOTA}, our method achieves superior segmentation accuracy on most of anatomical structures compared to other one-shot based methods.

\begin{table}[t]
    \footnotesize
    \centering
    \setlength{\tabcolsep}{6.0pt}
    \setlength{\extrarowheight}{1.5pt}
    \caption{Comparison with the state-of-the-art on CANDI. 
    The performance of fully supervised segmentation is also provided as an upper bound. 
    The Dice score (\%) is employed, and Mean(std) denotes the standard deviations.
    Min and Max denote the minimum and maximum Dice scores in the test set, respectively.}
    \vspace{-1em}
    \begin{tabular}{|l|ccc|}
    \hline
        Method & Mean(std) & Min & Max \\ \hline \hline
         Supervised learning & 88.3(1.7) & 83.5 & 90.3 \\ \hline
        VoxelMorph (Balakrishnan 2019)  & 76.0(9.7) & 61.7 & 80.1 \\ 
        DataAug (Zhao 2019) & 80.4(4.3) & 73.8 & 84.0 \\ 
        LT-Net (Wang 2020) & 82.3(2.5) &75.6 & 84.2\\ 
        Ours &\textbf{85.1}(\textbf{1.9}) & \textbf{80.2} & \textbf{87.8}  \\ \hline
    \end{tabular}
    \label{tab:SOTA CANDI}
\vspace{-0.5em}
\end{table}

\subsection{Ablation Study}
\subsubsection{Effectiveness of our VAEs}
\begin{table}[t]
    \footnotesize
    \centering
    \setlength{\tabcolsep}{2.0pt}
    \setlength{\extrarowheight}{1.5pt}
    \caption{Ablation study on different types of data augmentation.
    Shape and Intensity denote that the shape and intensity deformations are from registration. VAE indicates that the deformations are generated from our VAEs.}
    \vspace{-1em}
    \begin{tabular}{|l|ccc|ccc|}
    \hline
        Method & Shape & Intensity & VAE & Mean(std) & Min & Max \\ \hline \hline
        \multirow{2}*{\shortstack{Registration based\\(VoxelMorph)}}& & & & \multirow{2}*{76.0(9.7)} & \multirow{2}*{61.7} & \multirow{2}*{80.1} \\ 
        & & & & & & \\ \hline
        \multirow{4}*{\shortstack{Segmentation \\ with data \\ augmentation}}& $\surd$ & & & 81.7(5.6) & 65.4 & 87.4 \\ 
        &$\surd$ & & $\surd$& 83.5(4.2) & 71.1 & \textbf{87.8} \\ 
        & $\surd$&$\surd$ & & 84.2(\textbf{1.7}) &79.7 & 86.5\\ 
        &$\surd$ &$\surd$ & $\surd$&\textbf{85.1}(1.9) & \textbf{80.2} & \textbf{87.8}  \\ \hline
    \end{tabular}
    \label{tab: different data augmentation CANDI}
\vspace{-1em}
\end{table}
To demonstrate the effectiveness of our VAEs, we compare four different types of data augmentation in Table \ref{tab: different data augmentation CANDI}. 
As simply applying intensity offsets to the atlas does not change the segmentation mask, synthesized images will have the same segmentation labels, thus leading to a trivial segmentation solution. 

As indicated in Table \ref{tab: different data augmentation CANDI}, compared with direct registration, data augmentation based segmentation methods achieve better segmentation accuracy. 
Note that all the augmentation methods learn the shape deformations similar to VoxelMorph.
Compared with the data augmentation methods using deformations from image registration, our VAEs can generate richer data for training a segmentation network, thus leading to better performance.
Moreover, we observe that intensity deformations make great contributions to segmentation performance and various intensity changes facilitate the generalization of our segmentation network.
In Table \ref{tab: different data augmentation CANDI}, we also notice that our network employing registered shape and intensity deformations achieves better performance than DataAug. This is because DataAug pre-trains a segmentation network with an l2 loss and does not employ the atlas in training the segmentation network. Thus, using the atlas for training segmentation networks is important.

\begin{table}[t]
    \footnotesize
    \centering
    \setlength{\tabcolsep}{7.3pt}
    \setlength{\extrarowheight}{1.5pt}
    \caption{Ablation study on different reconstruction losses in the shape VAE.}
    \vspace{-1em}
    \begin{tabular}{|l|ccc|}
    \hline
        Method & Mean(std) & Min & Max \\ \hline \hline
        $\mL_S^{d}$  & 81.3 (\textbf{2.8}) & \textbf{74.4} & 85.0 \\ 
        $\mL_S^{i}$  & 82.3(6.2) & 63.9 & 87.7 \\ 
        $\mL_S^{d}$ + $\mL_S^{i}$ & \textbf{83.5}(4.2) & 71.1 & \textbf{87.8} \\ \hline
    \end{tabular}
    \label{tab:ablation study on rec loss}
\vspace{-1.0em}
\end{table}

\subsubsection{Effectiveness of the Combined Reconstruction Loss}
To demonstrate the effectiveness of our combined reconstruction loss \ie, $\mL_S^{d}+\mL_S^{i}$, we train the shape VAEs with $\mL_S^{d}$, $\mL_S^{i}$ and $\mL_S^{d}+\mL_S^{i}$, respectively, and then apply them to augment data. To avoid the influence of the intensity augmentation, we do not use intensity augmentation and the segmentation results are reported in Table \ref{tab:ablation study on rec loss}.
As indicated by Table \ref{tab:ablation study on rec loss}, our combined reconstruction loss is more suitable for the shape deformation learning and generation. 

\subsubsection{Hyper-parameter $\beta$ in Eq.~\eqref{eq:1} and Eq.~\eqref{eq:2}, and $\sigma$ for sampling latent codes}
\begin{figure}[t]
\centering
\scalebox{1}[0.85]{\includegraphics[width=0.99\linewidth]{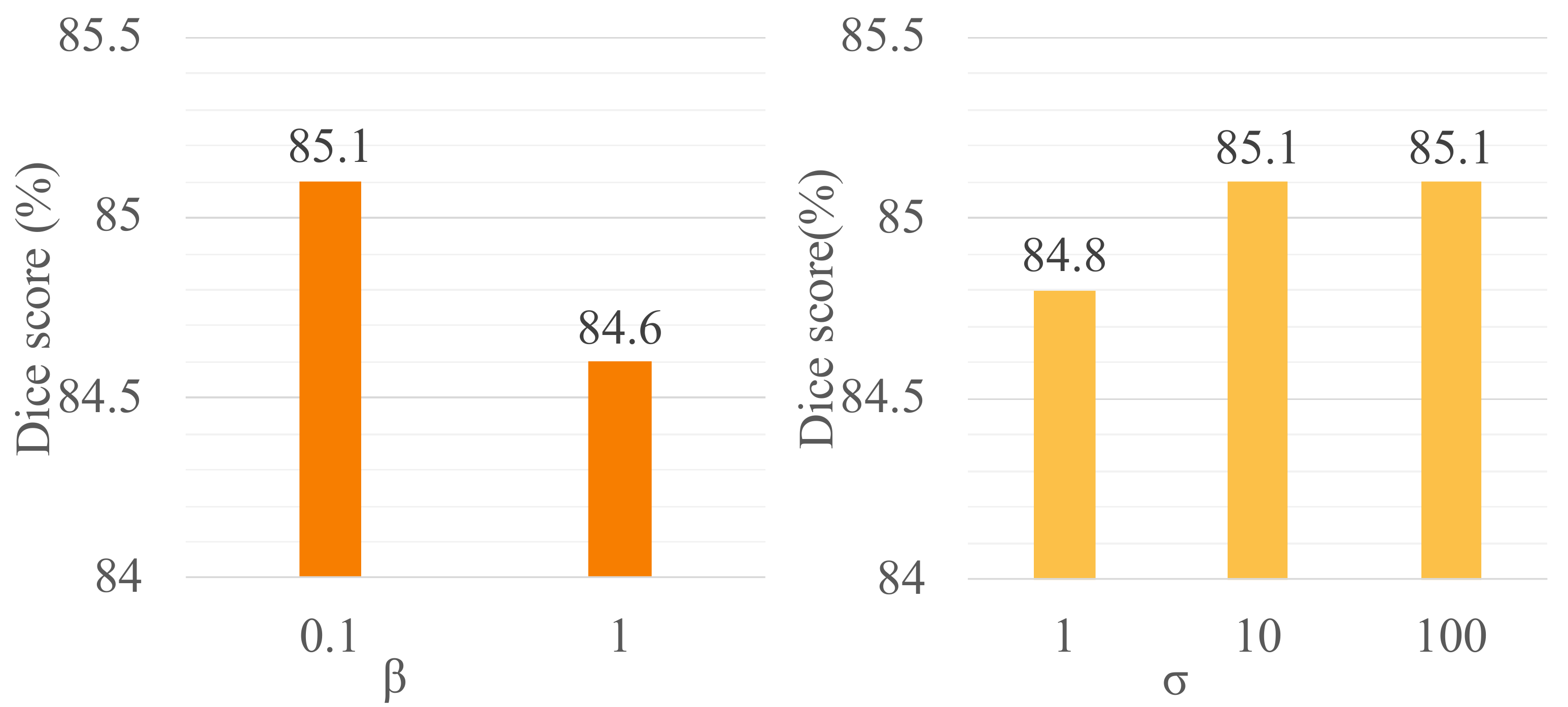}}
\vspace{-0.7em}
\caption{Ablation study on hyper-parameter $\beta$ and $\sigma$. 
$\beta$ controls the weight of the KL divergence and $\sigma$ is the standard deviation of a prior Gaussian distribution $\mathcal{N}(0,\sigma)$ in VAEs.
}
\vspace{-1.0em}
\label{Fig: ablation hyper parameter}
\end{figure}

As aforementioned, a small $\beta$ introduces more diversity into the generated deformations, thus improving the segmentation performance. 
Figure \ref{Fig: ablation hyper parameter} manifests that using a small $\beta$, we achieve better segmentation accuracy. Thus, in all the experiments, $\beta$ is set to $0.1$.
Furthermore, as illustrated in Fig. \ref{Fig: ablation hyper parameter}, the segmentation performance degrades when the standard deviation $\sigma$ for sampling latent codes is set to $1$. 
This is because we employ a small $\beta$ to enforce the KL divergence during training and the latent vector space would deviate from the standard normal distribution. Thus, we use a larger $\sigma$ to sample latent codes.
Figure \ref{Fig: ablation hyper parameter} shows the segmentation accuracy is similar when $\sigma$ is set to $10$ and $100$. Thus, $\sigma$ is set to $10$ for all the experiments.

\begin{table*}[t]
    \footnotesize
    \centering
    \setlength{\tabcolsep}{7.3pt}
    \setlength{\extrarowheight}{1.5pt}
    \caption{Comparison with the state-of-the-art methods on our newly proposed ABIDE benchmark.}
    \vspace{-1em}
    \begin{tabular}{|l|ccc|ccc|}
    \hline
        \multirow{2}*{Method} & \multicolumn{3}{c|}{Seen} & \multicolumn{3}{c|}{Unseen} \\ \cline{2-7}
        & Mean(std) & Min & Max & Mean(std) & Min & Max \\ \hline \hline
         Supervised learning & 87.6(2.7) & 79.3 & 91.1 &85.9(1.7) &81.3 & 87.5 \\ \hline
        VoxelMorph  & 70.3(11.6) & 33.1 & 82.5 &62.9(13.2) &32.3 &79.6  \\ 
        DataAug  & 69.6(9.02) & 39.7 & 80.4 & 64.3(9.9) & 35.0 &77.2\\ 
        Ours &\textbf{76.7(7.4)} & \textbf{53.2} & \textbf{86.5}  & \textbf{74.8(6.6)} & \textbf{54.1} & \textbf{83.3} \\ \hline
    \end{tabular}
    \label{tab:SOTA ABIDE}
\vspace{-1em}
\end{table*}
 
\subsection{Our Proposed ABIDE Benchmark}
\begin{figure}[t]
\centering
\scalebox{1}[1]{\includegraphics[width=0.99\linewidth]{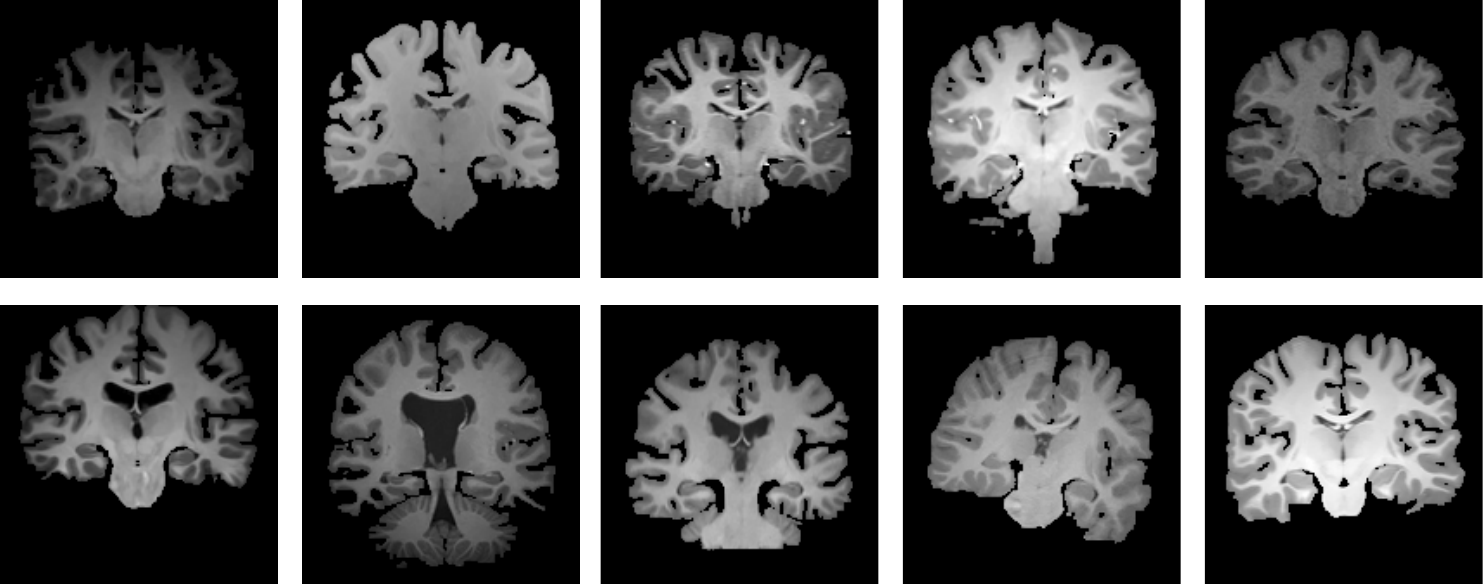}}
\vspace{-0.5em}
\caption{Illustration of significant variances in our ABIDE benchmark. The $96$-th slices of ten 3D MRI images are shown. (Top row: images from seen datasets; Bottom row: images from unseen datasets.) More images are shown in supplementary materials.
}\vspace{-1.0em}
\label{Fig:benchmark}
\end{figure}

Since the MRI images in CANDI are collected from only one source, the variances (including shape and intensity) mainly come from different individuals. 
However, different MRI machines and operations may also lead to variations.
Therefore, to validate the robustness of our method, we propose a new standard segmentation benchmark, called ABIDE benchmark, as visible in Fig.~\ref{Fig:benchmark}.

We sample T1-weighted MRI images from the autism brain imaging data exchange (ABIDE) database \cite{di2014autism}, which are collected from $17$ international sites.
We sample $190$ images from ten imaging sources and split them into $100$, $30$, $60$ volumes for training, validation and testing, respectively. These testing images form a \emph{seen} test set.
As suggested by \citet{DBLP:journals/tmi/BalakrishnanZSG19}, the most similar image to the average volume is selected as the atlas.
We also sample $60$ images from the rest imaging sources as an \emph{unseen} test set.
All the volumes are resampled into a $256 \times 256 \times 256$ with 1mm isotropic voxels and then cropped to $160 \times 160 \times 192$.
$28$ anatomical regions are annotated by FreeSurfer \cite{fischl2012freesurfer}.

As our benchmark contains images from multiple sites and includes an unseen test setting, it is more challenging and is also able to evaluate the robustness of a method. 

We compare our method with VoxelMorph\cite{DBLP:journals/tmi/BalakrishnanZSG19} and \D{DataAug\cite{DBLP:conf/cvpr/ZhaoBDGD19} 
}
in Table \ref{tab:SOTA ABIDE}. The performance of the segmentation network trained with full supervision is also reported.
Compared with the other two methods, we achieve superior performance on the seen and unseen datasets, demonstrating the effectiveness of our data augmentation method.
In addition, our performance only degrades 1.9\% on the unseen test dataset while the performance of the competing methods decreases more than 5\%. This demonstrates that our method achieves a better generalization ability with the help of our generated various deformations.

\vspace{-1.0em}
\section{Conclusion}
In this paper, we propose a 3D VAE based data augmentation scheme to generate realistic and diverse training samples for one-shot medical image segmentation.
We present a shape deformation VAE and an intensity deformation VAE to learn the distributions of the deformations of unlabeled real images with respect to an atlas one.
With the help of our learned VAEs, we can generate various deformations rather than solely combining existing deformations from unlabeled data, thus significantly enriching training data for segmentation.
To evaluate the segmentation performance on unseen imaging sources, we collect and annotate MRI data from different sources and construct a new benchmark. It provides us a standard public testbed for one-shot or few-shot based medical image segmentation methods. Extensive experiments demonstrate that our method outperforms the state-of-the-art on seen and unseen datasets.

\bibliography{main.bib}
\end{document}